\documentclass{article}

\newtheorem{fed}{\textbf{Definition}}[section]
\newtheorem{thm}[fed]{\textbf{Theorem}}
\newtheorem{lemma}[fed]{\textbf{Lemma}}

\newtheorem{cor}[fed]{\textbf{Corollary}}
\usepackage{amssymb,bbm,graphicx,epsfig,psfrag,epic,eepic,latexsym}
\usepackage{amsmath}
\usepackage{mathrsfs}
\usepackage[dvips]{color}

\begin{document}
\title{Helium and Hamiltonian delay equations}
\author{Urs Frauenfelder}
\maketitle

\begin{abstract}
In this paper we study two electrons on a line on the same side of the nucleus which interact with each other by their mean value. We prove that there exists a unique periodic orbit and examine for which charges the two orbits of the electrons intersect. 
\end{abstract}

\section{Introduction}

In the Helium atom two each other mutually repulsing electrons are attracted by a positive nucleus. This is a special case of a three-body problem. Due to the interaction between the two electrons the system is not integrable and shows chaotic features. This happens already if the two electrons are confined to a line, see \cite{wintgen-richter-tanner, tanner-richter-rost}. An interesting periodic orbit was detected numerically by Wintgen, Richter, and Tanner in \cite{wintgen-richter-tanner}. In this case both electron lie on the same side of the nucleus. The outer electron is almost stationary where the inner electron bounces back and forth to the nucleus. Since the outer electron is almost stationary this periodic orbit is referred to as the \emph{frozen-planet configuration}.
\bigskip

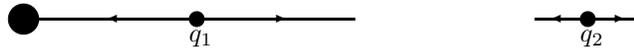
\setlength{\unitlength}{1mm}
\begin{picture}(1,1)
 \put(0,0){\circle*{4}}
 \put(2,0){\line(1,0){9}}
 \put(22,0){\vector(-1,0){11}}
 \put(23,0){\circle*{2}}
 \put(24,0){\vector(1,0){11}}
 \put(35,0){\line(1,0){9}}
 \put(68,0){\line(1,0){2}}
 \put(74,0){\vector(-1,0){4}}
 \put(75,0){\circle*{2}}
 \put(76,0){\vector(1,0){4}}
 \put(80,0){\line(1,0){2}}
 \put(22,-3){$q_1$}
 \put(74,-3){$q_2$}
\end{picture}
\bigskip
\bigskip

It is tempting to imagine that there is a Floer homology in which the frozen-planet configuration appears as a generator. 
The construction of such a Floer homology however is out of reach for this paper. This is due to the fact that the corresponding energy hypersurface is \emph{noncompact}. There are two reasons for that. One is that an electron can escape to infinity corresponding to ionization. The other reason are collisions. While two-body collisions can always be regularized \cite{levi-civita, moser} triple collisions provide a lot of difficulty. Of course in the frozen-planet configuration no triple collisions occur. However, other orbits might end up in a triple collision. On the other hand there are interesting connections with this question to current research. Due to the discovery of tentacular Hamiltonians Pasquotto, Vandervorst and Wi\'sniewska managed to construct a Floer homology for a class of noncompact hypersurfaces \cite{pasquotto-vandervorst-wisniewska}. The structure of the singularities at triple collisions were studied by McGehee \cite{mcgehee}. There are interesting relations of these singularities to b-symplectic geometry \cite{delshams-kiesenhofer-miranda}.

With the goal in mind to construct a Floer homology for this problem in the future and motivated by the joint work of the author with Albers, Schlenk, and Weber \cite{albers-frauenfelder-schlenk, frauenfelder-weber} on Hamiltonian delay equations we consider in this paper a model for the Helium atom where only the mean electrons interact. For a mean electron interaction the problem is not local and one therefore obtains a Hamiltonian delay equation. In the wave theoretical approach to quantum mechanics one often does not use local interactions between the electrons but averages about their probability distributions, see \cite{bethe-salpeter}. The author of this article does not know if such an approach is already used on the semiclassical side, namely the search for periodic orbits. 

In the actual Helium atom there are two protons in the nucleus and therefore the charge of the nucleus is two times the one of an electron. Here we more generally consider a nucleus of charge $\mu>1$. It is clear that for charge $\mu \leq 1$ there
cannot be any periodic orbits since in this case the outer electron is repelled by the inner electron more strongly than attracted by the nucleus so that it escapes to infinity and the atom ionizes. Our first main result is the following.
\\ \\
\textbf{Theorem\,A: } \emph{For every $\mu>1$ there is a unique simple unparametrized periodic orbit for mean electron interactions. }
\\ \\
The precise version of Theorem\,A is stated in Theorem~\ref{main1}. Since only the mean electrons interact the outer electron is actually frozen, namely constant, and not just almost frozen as in the case of local interactions. Moreover, 
the orbits of the two electrons might intersect, only the mean electrons are not allowed to intersect. The answer for which 
charges $\mu$ the two orbits intersect is related to the lemniscatic constant, introduced by Gau\ss. As $\pi$ denotes half the arc length of a circle its variation $\varpi$ denotes half the arclength of the lemniscate of Bernoulli. Where $\pi$ can be defined by the integral 
$$\pi=2\int_0^1 \frac{1}{\sqrt{1-t^2}}dt$$
the lemniscatic constant can be defined by
$$\varpi=2\int_0^1 \frac{1}{\sqrt{1-t^4}}dt.$$ 
It is also related to the Gamma function via
$$\varpi=\frac{\Gamma\big(\frac{1}{4}\big)^2}{\sqrt{8\pi}}\cong 2,62.$$
Our second theorem is
\\ \\
\textbf{Theorem\,B: } \emph{The two orbits intersect if and only if
$\mu\geq\big(\tfrac{3\pi}{3\pi-\varpi^2}\big)^2 \cong 13,69$.}
\\ \\
In particular, for the actual Helium $\mu=2$ and therefore in this case the two orbits do not intersect. 

\section{Variational approach to periodic orbits}

In this section we explain how in the classical set-up where the electrons act instantaneously on each other 
periodic orbits can be detected as critical points of a Lagrange multiplier action functional used by Rabinowitz
in his pioneering work on the applicability of global methods to Hamiltonian systems, see \cite{rabinowitz}. For compact
hypersurfaces a Floer homology for this action functional was constructed in \cite{cieliebak-frauenfelder} and its extension to noncompact hypersurfaces is a topic of active research \cite{pasquotto-vandervorst-wisniewska}. 

The configuration space for two electrons on a line on the same side of a nucleus at the origin is
$$Q=\big\{(q_1,q_2) \in \mathbb{R}^2: q_2>q_1>0\big\}.$$
The phase space is the cotangent bundle of the configuration space
$$T^*Q=Q \times \mathbb{R}^2.$$
The Hamiltonian is
$$H \colon T^*Q \to \mathbb{R}, \quad (q,p) \mapsto \frac{1}{2}|p|^2-\frac{\mu}{q_1}-\frac{\mu}{q_2}+\frac{1}{q_2-q_1}+1.$$
Here $\mu>1$ is the charge of the nucleus which for the actual Helium is $\mu=2$. We added one to the Hamiltonian to consider solutions on the level set $H^{-1}(0)$. On $T^*Q$ we have the Liouville one-form
$$\lambda=p_1 dq_1+p_2dq_2$$
whose exterior derivative is the standard symplectic form
$$\omega=dp_1 \wedge dq_1+dp_2\wedge dq_2.$$
The Hamiltonian equation of $H$ with respect to $\omega$ are then 
\begin{equation}\label{ham}
\left\{\begin{array}{c}
q'_1(t)=p_1(t)\\
q'_2(t)=p_2(t)\\
p'_1(t)=-\frac{\mu}{q_1(t)^2}-\frac{1}{(q_2(t)-q_1(t))^2}\\
p'_2(t)=-\frac{\mu}{q_2(t)^2}+\frac{1}{(q_2(t)-q_1(t))^2}.
\end{array}\right.
\end{equation}
If $S^1=\mathbb{R}/\mathbb{Z}$ is the circle we abbreviate
$$\mathcal{L}=C^\infty(S^1, T^*Q)$$
the space of free loops in the configuration space $T^*Q$. 
Rabinowitz action functional is defined as
$$\mathcal{A} \colon \mathcal{L}\times (0,\infty) \to \mathbb{R}, \quad (z,v) \mapsto \int z^*\lambda -\eta \int_0^1 H(z)dt$$
Its critical points are solutions of the problem
\begin{equation}\label{crit0}
\left\{\begin{array}{c}
q'_1(t)=\eta p_1(t)\\
q'_2(t)=\eta p_2(t)\\
p'_1(t)=-\frac{\mu\eta}{q_1(t)^2}-\frac{\eta}{(q_2(t)-q_1(t))^2}\\
p'_2(t)=-\frac{\mu\eta}{q_2(t)^2}+\frac{\eta}{(q_2(t)-q_1(t))^2}\\
H(q,p)=0.
\end{array}\right.
\end{equation}
After changing time by $t \mapsto \tfrac{t}{\eta}$ these are periodic solutions of (\ref{ham}) of period $\eta$ subject to the energy constraint $H(q,p)=0$.

\section{Mean electron interactions}

In this section we explain how we obtain mean electron interactions by considering a delayed Rabinowitz action functional.
For $q \in C^\infty(S^1,\mathbb{R})$ denote its average by
$$\overline{q}=\int_0^1 q(t)dt.$$
Abbreviate
$$\mathcal{L}=\Big\{(q,p) \in C^\infty\big(S^1,T^*(0,\infty)^2\big): \overline{q}_2>\overline{q}_1\Big\}$$
and define the delayed Rabinowitz action functional
$$\mathcal{A}\colon \mathcal{L} \times (0,\infty) \to \mathbb{R}$$
for $z=(q,p)=(q_1,q_2,p_1,p_2) \in \mathcal{L}$ and $\eta \in (0,\infty)$ by
$$\mathcal{A}(z,\eta)=\int z^*\lambda-\eta\Bigg(\int_0^1 \bigg(\frac{1}{2}|p|^2-\frac{\mu}{q_1}-\frac{\mu}{q_2}\bigg)dt+
\frac{1}{\overline{q}_2-\overline{q}_1}+1\Bigg).$$
Critical points of Rabinowitz action function $\mathcal{A}$ are tuples
satisfying the following Hamiltonian delay equation 
\begin{equation}\label{crit1}
\left\{\begin{array}{c}
q'_1(t)=\eta p_1(t)\\
q'_2(t)=\eta p_2(t)\\
p'_1(t)=-\frac{\mu\eta}{q_1(t)^2}-\frac{\eta}{(\overline{q}_2-\overline{q}_1)^2}\\
p'_2(t)=-\frac{\mu\eta}{q_2(t)^2}+\frac{\eta}{(\overline{q}_2-\overline{q}_1)^2}\\
\int_0^1 \bigg(\frac{1}{2}|p|^2-\frac{\mu}{q_1}-\frac{\mu}{q_2}\bigg)dt+
\frac{1}{\overline{q}_2-\overline{q}_1}+1=0.
\end{array}\right.
\end{equation}
Note that the momenta $p$ are uniquely determined by the period $\eta$ and the velocities $q'$ so that the problem (\ref{crit1}) is equivalent to the following second order delay problem
\begin{equation}\label{crit2}
\left\{\begin{array}{c}
q''_1(t)=-\frac{\mu\eta^2}{q_1(t)^2}-\frac{\eta^2}{(\overline{q}_2-\overline{q}_1)^2}\\
q''_2(t)=-\frac{\mu\eta^2}{q_2(t)^2}+\frac{\eta^2}{(\overline{q}_2-\overline{q}_1)^2}\\
\int_0^1 \Big(\tfrac{(q'_1)^2}{2\eta^2}+\tfrac{(q'_2)^2}{2\eta^2}-\frac{\mu}{q_1}-\frac{\mu}{q_2}\Big)dt+
\frac{1}{\overline{q}_2-\overline{q}_1}+1=0.
\end{array}\right.
\end{equation}
From the first equation we conclude that
$$q_1''(t)<0, \quad \forall\,\,t \in S^1$$
implying that there are no periodic solutions. In particular, there are no critical points. This a bit disappointing situation however can be remedied if one allows collisions of the first electron with the kernel. 

\subsection{The regularized problem}

In celestial mechanics it is known very well that one can always regularize two-body collisions. There are different procedures how this can be carried out, see for instance \cite{levi-civita, moser}. If the particle moves on a half line
it just bounces back on the half line after the collision. We do not address here the question how to regularize the action functional. We just explain how to regularize its critical points. We plan to address the issue of regularization of the action functional in a future paper so that the regularized critical points can be interpreted as critical points of the regularized action functional. 

We allow collisions of the first electron but not of the second one. Therefore the second equation in (\ref{crit2}) still remains valid for all times. In the following lemma we show that this implies that $q_2$ is necessarily constant.
\begin{lemma}\label{constant} 
Assume that $q_2 \in C^\infty\big(S^1,(0,\infty)\big)$ solves the second equation in (\ref{crit2}). Then $q_2$ is constant. 
\end{lemma}
\textbf{Proof: } Since the circle is compact there exists $t_0 \in S^1$ such that
$$q_2(t_0):=\mathrm{min}\{q_2(t): t \in S^1\}.$$
In particular, we have
$$q_2''(t_0) \geq 0.$$
Since $q_2$ attains at $t_0$ its absolute minimum we conclude from the second equation in (\ref{crit2}) that
$$q_2''(t) \geq 0, \quad \forall\,\,t \in S^1.$$
Since $q_2$ is periodic this implies that
$$q_2''(t)=0, \quad \forall\,\,t \in S^1.$$
In particular, the velocity of $q_2$ is constant. Using again that $q_2$ is periodic we obtain that
$$q_2'(t)=0, \quad \forall\,\,t \in S^1$$
so that $q_2$ is indeed constant. This finishes the proof of the Lemma. \hfill $\square$
\\ \\
In view of Lemma~\ref{constant} we have
\begin{equation}\label{qzwei}
q_2(t)=\overline{q}_2, \quad \forall\,\,t \in S^1
\end{equation}
such that the second equation in (\ref{crit2}) becomes
$$0=-\frac{\mu}{\overline{q}_2^2}+\frac{1}{(\overline{q}_2-\overline{q}_1)^2}$$
which we can rewrite as
$$\mu(\overline{q}_2-\overline{q}_1)^2=\overline{q}_2^2.$$
This is equivalent to the homogeneous equation of degree two in $\overline{q}_1$ and $\overline{q}_2$
$$(\mu-1)\overline{q}_2^2-2\mu\overline{q}_1 \overline{q}_2+\mu\overline{q}_1^2=0.$$
In particular, we obtain for $\overline{q}_2$
$$\overline{q}_2=\frac{2\mu\overline{q}_1\pm\sqrt{4\mu^2\overline{q}_1^2-4\mu(\mu-1)\overline{q}_1^2}}{2(\mu-1)}=
\frac{2\mu\overline{q}_1\pm\sqrt{4\mu\overline{q}_1^2}}{2(\mu-1)}=\frac{\mu \pm \sqrt{\mu}}{\mu-1}\overline{q}_1$$
Since $\overline{q}_2>\overline{q}_1>0$ we get
\begin{equation}\label{barqzwei}
\overline{q}_2=\frac{\mu +\sqrt{\mu}}{\mu-1}\overline{q}_1
\end{equation}
In particular, we see from (\ref{qzwei}) and (\ref{barqzwei}) that $q_2$ is completely determined by the average of $q_1$.
From (\ref{barqzwei}) we obtain
$$\overline{q}_2-\overline{q}_1=\bigg(\frac{\mu+\sqrt{\mu}}{\mu-1}-1\bigg)\overline{q_1}=\frac{1+\sqrt{\mu}}{\mu-1}
\overline{q}_1$$
Abbreviating 
\begin{equation}\label{defgam}
\gamma:=\gamma(\mu):=\frac{1+\sqrt{\mu}}{\mu-1}
\end{equation}
we can replace the first equation in (\ref{crit2}) by
\begin{equation}\label{qeins}
q_1''(t)=-\frac{\mu \eta^2}{q_1(t)^2}-\frac{\eta^2}{\gamma^2 \overline{q}_1^2}.
\end{equation}
For later reference we note that
\begin{equation}\label{gameq}
\sqrt{\mu}\gamma=\frac{\mu+\sqrt{\mu}}{\mu-1}=\gamma+1
\end{equation}
so that (\ref{barqzwei}) can be equivalently rewritten as
\begin{equation}\label{barqzwei2}
\overline{q}_2=(\gamma+1)\overline{q}_1.
\end{equation}
From (\ref{gameq}) we see as well that we can express $\mu$ with the help of $\gamma$ by
\begin{equation}\label{mueq}
\mu=\frac{(\gamma+1)^2}{\gamma^2}.
\end{equation}
In particular,
$$\gamma \colon (1,\infty) \to (0,\infty)$$
is an orientation reversing diffeomorphism.
\\ \\
Physically the ODE (\ref{qeins}) means that the first electron is attracted by the protons in the nucleus according to Coulomb's law and is subject to an additional constant force depending on its own mean value. An integral for this system is
\begin{equation}\label{energy}
\kappa=\frac{1}{2\eta^2}(q_1'(t))^2-\frac{\mu}{q_1(t)}+\frac{q_1(t)}{\gamma^2 \overline{q}_1^2}.
\end{equation}
Suppose that at time $t=0$ the first electron is at position $q_1^{\mathrm{max}}$ with zero velocity. Then the electron gets accelerated towards the nucleus until it finally collides with infinite velocity with the nucleus. Note that equation (\ref{qeins}) is invariant under time reversal. Hence we just let the movie run backwards. The electron jumps out of the nucleus with infinite velocity and decelerates until it is back at $q_1^{\mathrm{max}}$ with zero velocity. As the notation suggests the point $q_1^{\mathrm{\max}}$ actually corresponds to the maximum of $q_1$. Up to time shift all regularized solutions of (\ref{qeins}) are of this form. In particular, all solutions are periodic. In the following we parametrize our solution such that at time $t=0$ it attains its maximum $q_1^{\mathrm{max}}$ and moreover we assume that the periodic orbit is simple so that it has precisely one collision. A general periodic orbit is then a multiple cover of a simple one.  

Since at the maximal point the velocity vanishes we obtain from (\ref{energy})
$$\kappa=-\frac{\mu}{q_1^{\mathrm{max}}}+\frac{q_1^{\mathrm{max}}}{\gamma^2 \overline{q}_1^2}.$$
In particular, this gives rise to the quadratic equation for $q_1^{\mathrm{max}}$
$$\big(q_1^{\mathrm{max}}\big)^2-\gamma^2 \kappa \overline{q}_1^2 q_1^{\mathrm{max}}-\mu \gamma^2 \overline{q}_1^2=0$$
with solutions
$$q_1^{\mathrm{max}}=\frac{\gamma^2 \kappa \overline{q}_1^2\pm \sqrt{\gamma^4\kappa^2\overline{q}_1^4+4\mu\gamma^2 \overline{q}_1^2}}{2}.$$
Since 
$$q_1^{\mathrm{max}}>0$$
we conclude that
\begin{equation}\label{maximum}
q_1^{\mathrm{max}}=\frac{\gamma^2 \kappa \overline{q}_1^2 + \sqrt{\gamma^4\kappa^2\overline{q}_1^4+4\mu\gamma^2 \overline{q}_1^2}}{2}=\frac{\gamma^2 \kappa \overline{q}_1^2 + \sqrt{\gamma^4\kappa^2\overline{q}_1^4+4(\gamma+1)^2\overline{q}_1^2}}{2}.
\end{equation}
In particular, we see from (\ref{maximum}) that $q_1^{\mathrm{max}}$ is uniquely determined by the tuple
$(\kappa,\overline{q}_1)$. We can characterize $q_1$ to be the unique solution of the second order ODE (\ref{qeins})
satisfying the initial condition
$$q_1(0)=q_1^{\mathrm{max}}, \qquad q_1'(0)=0.$$
The ODE (\ref{qeins}) depends on the period $\eta>0$. The requirement
$$q_1\big(\tfrac{1}{2}\big)=0$$
determines $\eta$ uniquely. Indeed, from (\ref{energy}) we have
\begin{equation}\label{ableitung}
q_1'(t)=\eta \sqrt{2\kappa +\frac{2\mu}{q_1(t)}-\frac{2q_1(t)}{\gamma^2 \overline{q}_1^2}}.
\end{equation}
Therefore, we obtain
$$\frac{1}{2}=\int_0^{1/2} dt=\int_0^{q_1^{\mathrm{max}}}\frac{1}{\eta \sqrt{2\kappa +\frac{2\mu}{q_1}-\frac{2q_1}{\gamma^2 \overline{q}_1^2}}}dq_1$$ 
implying that
\begin{equation}\label{period}
\eta=\sqrt{2}\int_0^{q_1^{\mathrm{max}}}\frac{1}{\sqrt{\kappa +\frac{\mu}{q_1}-\frac{q_1}{\gamma^2 \overline{q}_1^2}}}dq_1.
\end{equation}
Using (\ref{ableitung}) once more we obtain
$$\overline{q}_1=2\int_0^{1/2} q_1(t) dt=2\int_0^{q_1^{\mathrm{max}}}\frac{q_1}{\eta \sqrt{2\kappa +\frac{2\mu}{q_1}-\frac{2q_1}{\gamma^2 \overline{q}_1^2}}}dq_1$$
so that using (\ref{period}) we end up with the following fixpoint problem for $\overline{q}_1$
\begin{equation}\label{fixpunkt}
\overline{q}_1=\frac{\int_0^{q_1^{\mathrm{max}}}\frac{q_1}{\sqrt{\kappa +\frac{\mu}{q_1}-\frac{q_1}{\gamma^2 \overline{q}_1^2}}}dq_1}{\int_0^{q_1^{\mathrm{max}}}\frac{1}{\sqrt{\kappa +\frac{\mu}{q_1}-\frac{q_1}{\gamma^2 \overline{q}_1^2}}}dq_1}.
\end{equation}
By the discussion above a tuple $(\kappa,\overline{q}_1) \in \mathbb{R}\times (0,\infty)$ satisfying the fixpoint problem (\ref{fixpunkt}) is up to time-shift in one-to-one correspondence with a periodic solution of (\ref{qeins}) having one collision. 

The ODE (\ref{qeins}) corresponds to the first equation in (\ref{crit2}). We next rewrite the energy equation namely the third equation in (\ref{crit2}). Using that $q_2$ is constant and taking advantage of (\ref{energy}) and (\ref{barqzwei2}) we obtain 
\begin{eqnarray*}
0&=&\int_0^1\bigg(\frac{1}{2}p_1^2+\frac{1}{2}p_2^2-\frac{\mu}{q_1}-\frac{\mu}{q_2}\bigg)dt+\frac{1}{\overline{q}_2-\overline{q}_1}
+1\\
&=&\int_0^1\bigg(\frac{1}{2}p_1^2-\frac{\mu}{q_1}\bigg)dt-\frac{\mu}{\overline{q}_2}+\frac{1}{\overline{q}_2-\overline{q}_1}
+1\\
&=&\int_0^1\bigg(\kappa-\frac{q_1}{\gamma^2 \overline{q}^2_1}\bigg)dt-\frac{\mu}{(\gamma+1)\overline{q}_1}+\frac{1}{\gamma\overline{q}_1}
+1\\
&=&\kappa-\frac{1}{\gamma^2 \overline{q}_1^2}\int_0^1 q_1dt-\frac{\mu}{(\gamma+1)\overline{q}_1}+\frac{1}{\gamma\overline{q}_1}
+1\\
&=&\kappa-\frac{1}{\gamma^2 \overline{q}_1}-\frac{\mu}{(\gamma+1)\overline{q}_1}+\frac{1}{\gamma\overline{q}_1}
+1
\end{eqnarray*}
Using (\ref{gameq}) we can simplify
\begin{eqnarray*}
\frac{1}{\gamma}-\frac{\mu}{\gamma+1}-\frac{1}{\gamma^2}&=&
\frac{\gamma(\gamma+1)-\mu\gamma^2-(\gamma+1)}{\gamma^2(\gamma+1)}\\
&=&\frac{\gamma(\gamma+1)-(\gamma+1)^2-(\gamma+1)}{\gamma^2(\gamma+1)}\\
&=&\frac{\gamma-(\gamma+1)-1}{\gamma^2}\\
&=&-\frac{2}{\gamma^2}
\end{eqnarray*}
so that we obtain
\begin{equation}\label{kapeq}
\kappa=\frac{2}{\gamma^2 \overline{q}_1}-1.
\end{equation}
Plugging (\ref{kapeq}) into (\ref{fixpunkt}) we obtain taking advantage of (\ref{gameq}) as well
\begin{eqnarray}\label{fixpoint2}
\overline{q}_1&=&\frac{\int_0^{q_1^{\mathrm{max}}}\frac{q_1}{\sqrt{\frac{2}{\gamma^2 \overline{q}_1}-1 +\frac{\mu}{q_1}-\frac{q_1}{\gamma^2 \overline{q}_1^2}}}dq_1}{\int_0^{q_1^{\mathrm{max}}}\frac{1}{\sqrt{\frac{2}{\gamma^2 \overline{q}_1}-1+\frac{\mu}{q_1}-\frac{q_1}{\gamma^2 \overline{q}_1^2}}}dq_1}\\ \nonumber
&=&\frac{\int_0^{q_1^{\mathrm{max}}}\frac{q_1^{3/2}}{\sqrt{2\overline{q}_1q_1-\gamma^2\overline{q}_1^2q_1+\mu\gamma^2 \overline{q}_1^2-q_1^2}}dq_1}{\int_0^{q_1^{\mathrm{max}}}\frac{q_1^{1/2}}{\sqrt{2\overline{q}_1q_1-\gamma^2\overline{q}_1^2q_1+\mu\gamma^2 \overline{q}_1^2-q_1^2}}dq_1}\\ \nonumber
&=&\frac{\int_0^{q_1^{\mathrm{max}}}\frac{q_1^{3/2}}{\sqrt{2\overline{q}_1q_1-\gamma^2\overline{q}_1^2q_1+(\gamma+1)^2 \overline{q}_1^2-q_1^2}}dq_1}{\int_0^{q_1^{\mathrm{max}}}\frac{q_1^{1/2}}{\sqrt{2\overline{q}_1q_1-\gamma^2\overline{q}_1^2q_1+(\gamma+1)^2 \overline{q}_1^2-q_1^2}}dq_1}.
\end{eqnarray}
If we plug in (\ref{kapeq}) into (\ref{maximum}) we get
\begin{eqnarray*}
q_1^{\mathrm{max}}&=&\frac{\gamma^2 \Big(\frac{2}{\gamma^2\overline{q}_1}-1\Big) \overline{q}_1^2 + \sqrt{\gamma^4\Big(\frac{2}{\gamma^2\overline{q}_1}-1\Big)^2\overline{q}_1^4+4(\gamma+1)^2 \overline{q}_1^2}}{2}\\
&=&\frac{(2-\gamma^2\overline{q}_1) \overline{q}_1 + \sqrt{(2-\gamma^2\overline{q}_1)^2\overline{q}_1^2+4(\gamma+1)^2 \overline{q}_1^2}}{2}\\
&=&\frac{\overline{q}_1}{2}\Big(2-\gamma^2\overline{q}_1+\sqrt{(2-\gamma^2\overline{q}_1)^2+4(\gamma+1)^2}\Big).
\end{eqnarray*}
We further set
$$q_1^{\mathrm{antimax}}
=\frac{\overline{q}_1}{2}\Big(2-\gamma^2\overline{q}_1-\sqrt{(2-\gamma^2\overline{q}_1)^2+4(\gamma+1)^2}\Big).
$$
While $q_1^{\mathrm{max}}$ corresponds to the maximal point of the trajectory $q_1$ the point $q_1^{\mathrm{antimax}}<0$ has no
physical interpretation. However, $q_1^{\mathrm{max}}$ and $q_1^{\mathrm{antimax}}$ are the roots of the quadratic polynomial in the denominator of the integrals on the righthand side of (\ref{fixpoint2}), namely
$$2\overline{q}_1q_1-\gamma^2\overline{q}_1^2q_1+(\gamma+1)^2 \overline{q}_1^2-q_1^2=\big(q_1^{\mathrm{max}}-q_1\big)\big(q_1-
q_1^{\mathrm{antimax}}\big).$$
Therefore we can write (\ref{fixpoint2}) more compactly
$$\overline{q}_1=\frac{\int_0^{q_1^{\mathrm{max}}}\frac{q_1^{3/2}}{\sqrt{(q_1^{\mathrm{max}}-q_1)(q_1-
q_1^{\mathrm{antimax}})}}dq_1}{\int_0^{q_1^{\mathrm{max}}}\frac{q_1^{1/2}}{\sqrt{(q_1^{\mathrm{max}}-q_1)(q_1-
q_1^{\mathrm{antimax}})}}dq_1}.$$
We change variables in the integrals by setting
$$r=\frac{q_1}{q_1^\mathrm{max}}$$
to get
\begin{equation}\label{fixpunkt3}
\overline{q}_1=\frac{(q_1^{\mathrm{max}})^2\int_0^1\frac{r^{3/2}}{\sqrt{(1-r)(q_1^{\mathrm{max}}r-
q_1^{\mathrm{antimax}})}}dr}{q_1^\mathrm{max}\int_0^1\frac{r^{1/2}}{\sqrt{(1-r)(q_1^{\mathrm{max}}r-
q_1^{\mathrm{antimax}})}}dr}=
\frac{q_1^{\mathrm{max}}\int_0^1\frac{r^{3/2}}{\sqrt{(1-r)(q_1^{\mathrm{max}}r-
q_1^{\mathrm{antimax}})}}dr}{\int_0^1\frac{r^{1/2}}{\sqrt{(1-r)(q_1^{\mathrm{max}}r-
q_1^{\mathrm{antimax}})}}dr}.
\end{equation}
We further introduce the functions
$$\sigma^+:=\sigma^+(\gamma,\overline{q}_1):=\frac{2q_1^\mathrm{max}}{\overline{q}_1}=2-\gamma^2\overline{q}_1+\sqrt{(2-\gamma^2\overline{q}_1)^2+4(\gamma+1)^2}$$
and
$$\sigma^-:=\sigma^-(\gamma,\overline{q}_1):=\frac{2q_1^\mathrm{antimax}}{\overline{q}_1}=2-\gamma^2\overline{q}_1-\sqrt{(2-\gamma^2\overline{q}_1)^2+4(\gamma+1)^2}.$$
With these notions we can reformulate (\ref{fixpunkt3}) to 
\begin{equation}\label{fixpunkt4}
2=\frac{\sigma^+\int_0^1\frac{r^{3/2}}{\sqrt{(1-r)(\sigma^+ r-
\sigma^-)}}dr}{\int_0^1\frac{r^{1/2}}{\sqrt{(1-r)(\sigma^+ r-\sigma^-)}}dr}.
\end{equation}
If we introduce the function
$$\mathcal{F}\colon (0,\infty)^2 \to (0,\infty), \quad
(\gamma,x) \mapsto 
\frac{\sigma^+(\gamma,x)\int_0^1\frac{r^{3/2}}{\sqrt{(1-r)(\sigma^+(\gamma,x) r-
\sigma^-(\gamma,x))}}dr}{\int_0^1\frac{r^{1/2}}{\sqrt{(1-r)(\sigma^+(\gamma,x) r-
\sigma^-(\gamma,x))}}dr}$$
we can simplify this to the compact expression
$$2=\mathcal{F}(\gamma,\overline{q}_1).$$
For fixed $\gamma>0$ we further abbreviate
$$\mathcal{F}_\gamma:=\mathcal{F}(\gamma,\cdot) \colon (0,\infty) \to (0,\infty).$$
In view of the discussion of this section the moduli space of simple regularized unparametrized solutions of (\ref{crit1})
for charge of the nucleus $\mu>1$ is in one-to-one correspondence with the set
$$\mathcal{M}(\mu):=\mathcal{F}_{\gamma(\mu)}^{-1}(2)$$
where $\gamma(\mu)$ is given by (\ref{gameq}). The following Theorem is the precise version of Theorem\,A from the introduction. 
\begin{thm}\label{main1}
For every $\mu>1$ the set $\mathcal{M}(\mu)$ consists of a unique point. 
\end{thm}
\section{Analysis of the function $\mathcal{F}$}

In this section we prove some useful properties of the function $\mathcal{F}$ and prove Theorem~\ref{main1}.

\begin{lemma}\label{func1}
For every $\gamma>0$ the function $\mathcal{F}_\gamma$ is strictly monotone decreasing.
\end{lemma}
\textbf{Proof: }We abbreviate $\sigma^+_\gamma=\sigma^+(\gamma,\cdot)$ and $\sigma^-_\gamma=\sigma^-(\gamma,\cdot)$.
We first check that both functions are strictly monotone decreasing. Indeed,
$$(\sigma^+_\gamma)'(x)=-\gamma^2-\frac{(2-\gamma^2 x)\gamma^2}{\sqrt{(2-\gamma^2 x)^2+4(\gamma+1)^2}}<-\gamma^2+\gamma^2=0$$
and similarly
$$(\sigma^-_\gamma)'(x)=-\gamma^2+\frac{(2-\gamma^2 x)\gamma^2}{\sqrt{(2-\gamma^2 x)^2+4(\gamma+1)^2}}<-\gamma^2+\gamma^2=0.$$
Since $\sigma^+>0$ and $\sigma^-<0$ we conclude that
$\tfrac{\sigma^-_\gamma}{\sigma^+_\gamma}$ is strictly monotone decreasing as well. Moreover,
$$\sigma^+\sigma^-=(2-\gamma^2 x)^2-(2-\gamma^2 x)^2-4(\gamma+1)^2=-4(\gamma+1)^2$$
is indepenent of $x$ and therefore $\sigma^+_\gamma \sigma^-_\gamma$ is constant. Hence writing
$$\mathcal{F}=\frac{\int_0^1\frac{r^{3/2}}{\sqrt{(1-r)\big(r-
\frac{\sigma^-}{\sigma^+}\big)}}dr}{\int_0^1\frac{r^{1/2}}{\sqrt{(1-r)\big((\sigma^+)^2 r-\sigma^-\sigma^+\big)}}dr}$$
we see that the numerator is strictly monotone decreasing and the denominator is strictly monotone increasing so that 
$\mathcal{F}_\gamma$ is strictly monotone decreasing. \hfill $\square$
\\ \\
For physical reasons the function $\mathcal{F}$ was only defined on the domain $(0,\infty)^2$. However, by the same formula
we actually obtain a smooth function on $\mathbb{R}^2$. In the following by abuse of notation we denote this extension
by the same letter. 
\begin{lemma}\label{func2} 
At the point $(-1,0)$ the function takes the value
$\mathcal{F}(-1,0)=\tfrac{8}{3}$
\end{lemma}
\textbf{Proof: } We have
$$\sigma^+(-1,0)=4,\qquad \sigma^-(-1,0)=0.$$
Therefore
\begin{eqnarray*}
\mathcal{F}(-1,0)&=&4\cdot\frac{\int_0^1\frac{r^{3/2}}{\sqrt{4r(1-r)}}dr}{\int_0^1\frac{r^{1/2}}{\sqrt{4r(1-r)}}dr}\\
&=&4\cdot\frac{\int_0^1\frac{r}{\sqrt{1-r}}dr}{\int_0^1\frac{1}{\sqrt{1-r}}dr}\\
&=&4\cdot \frac{-\frac{2\sqrt{1-r}(r+2)}{3}\Big|_0^1}{-2\sqrt{1-x}\big|_0^1}\\
&=&4\cdot \frac{\frac{4}{3}}{2}\\
&=&\frac{8}{3}.
\end{eqnarray*}
This finishes the proof of the lemma. \hfill $\square$
\\ \\
In the following lemma we fix the $x$-variable to be zero and set 
$$\mathcal{F}^0=\mathcal{F}(\cdot,0) \colon \mathbb{R} \to (0,\infty).$$
\begin{lemma}\label{func3}
For $\gamma \geq -1$ the function $\mathcal{F}^0$ is monotone increasing.
\end{lemma}
\textbf{Proof: } Abbreviate for $\gamma \in \mathbb{R}$
$$\rho_+(\gamma):=\sigma^+(\gamma,0)=2+2\sqrt{1+(\gamma+1)^2}$$
and
$$\rho_-(\gamma):=\sigma^-(\gamma,0)=2-2\sqrt{1+(\gamma+1)^2}$$
Note that
\begin{eqnarray*}
(\rho_-' \rho_+-\rho_+' \rho_-)(\gamma)&=&-\frac{2(\gamma+1)(2+2\sqrt{1+(\gamma+1)^2}}{\sqrt{1+(\gamma+1)^2}}\\
& &-\frac{2(\gamma+1)(2-2\sqrt{1+(\gamma+1)^2}}{\sqrt{1+(\gamma+1)^2}}\\
&=&-\frac{8(\gamma+1)}{\sqrt{1+(\gamma+1)^2}}\\
&\leq&0
\end{eqnarray*}
so that
$\tfrac{\rho_-}{\rho_+}$ is monotone decreasing. In particular, the function
$$\gamma \mapsto \int_0^1\frac{r^{1/2}}{\sqrt{(1-r)\Big(r-\frac{\rho_-(\gamma)}{\rho_+(\gamma)}\Big)}}dr$$
is monotone decreasing.
\\ \\
Note that since $\gamma \geq -1$
$$\rho_+'=-\rho_-' \geq 0$$
and moreover, $\rho_+$ is positive where $\rho_-$ is nonpositive. Therefore
for $r \in [0,1]$ we have
\begin{equation}\label{schreck1}
\bigg(\frac{r}{\rho_+^2}-\frac{\rho_-}{\rho_+^3}\bigg)'=-\frac{2r\rho_+'}{\rho_+^3}-\frac{\rho_-'}{\rho_+^3}+
\frac{3\rho_-\rho_+'}{\rho_+^4}\leq -\frac{2r\rho_+'}{\rho_+^3}-\frac{\rho_-'}{\rho_+^3}=(1-2r)\frac{\rho_+'}{\rho_+^3}.
\end{equation}
Consider the function
$$\mathcal{G}:=\int_0^1\frac{r^{3/2}}{\sqrt{(1-r)\Big(\frac{r}{\rho_+^2}-
\frac{\rho_-}{\rho_+^3}\Big)}}dr$$
Due to inequality (\ref{schreck1}) its derivative can be estimated as
\begin{eqnarray}\label{schreck2}
\mathcal{G}'&=&-\frac{1}{2}\int_0^1 \frac{r^{3/2}\Big(\frac{r}{\rho_+^2}-\frac{\rho_-}{\rho^3_+}\Big)'}{\sqrt{(1-r)\Big(\frac{r}{\rho^2_+}-\frac{\rho_-}{\rho^3_+}\Big)^3}}dr\\ \nonumber
&\geq&-\frac{1}{2}\int_0^1 \frac{r^{3/2}(1-2r)\frac{\rho_+'}{\rho_+^3}}{\sqrt{(1-r)\Big(\frac{r}{\rho^2_+}-\frac{\rho_-}{\rho^3_+}\Big)^3}}dr\\ \nonumber
&=&-\frac{\rho_+'}{2}\int_0^1\frac{1-2r}{\sqrt{(1-r)\Big(1-\frac{\rho_-}{\rho_+ r^{3/2}}\Big)}}dr
\end{eqnarray}
To estimate this further we set
$$a:=-\frac{\rho_-}{\rho_+} \geq 0.$$
Then
\begin{eqnarray}\label{schreck3}
\int_0^1\frac{1-2r}{\sqrt{(1-r)\Big(1+\frac{a}{r^{3/2}}\Big)}}dr&=&\int_0^{1/2}\frac{1-2r}{\sqrt{(1-r)\Big(1+\frac{a}{r^{3/2}}\Big)}}dr\\ \nonumber
& &+\int_{1/2}^1\frac{1-2r}{\sqrt{(1-r)\Big(1+\frac{a}{r^{3/2}}\Big)}}dr\\ \nonumber
&=&\int_0^{1/2}\frac{2s}{\sqrt{\Big(\frac{1}{2}+s\Big)\Big(1+\frac{a}{(\frac{1}{2}-s)^{3/2}}\Big)}}dr\\ \nonumber
& &-\int_0^{1/2}\frac{2s}{\sqrt{\Big(\frac{1}{2}-s\Big)\Big(1+\frac{a}{(\frac{1}{2}+s)^{3/2}}\Big)}}dr\\ \nonumber
&\leq&\int_0^{1/2}\frac{2s}{\sqrt{\Big(\frac{1}{2}-s\Big)\Big(1+\frac{a}{(\frac{1}{2}+s)^{3/2}}\Big)}}dr\\ \nonumber
& &-\int_0^{1/2}\frac{2s}{\sqrt{\Big(\frac{1}{2}-s\Big)\Big(1+\frac{a}{(\frac{1}{2}+s)^{3/2}}\Big)}}dr\\ \nonumber
&\leq&0.
\end{eqnarray}
Combining (\ref{schreck2}) and (\ref{schreck3}) we conclude that
$$\mathcal{G}' \geq 0$$
so that $\mathcal{G}$ is monotone increasing. We now write
\begin{eqnarray*}
\mathcal{F}^0&=&\frac{\rho_+\int_0^1\frac{r^{3/2}}{\sqrt{(1-r)(\rho_+r-
\rho_-)}}dr}{\int_0^1\frac{r^{1/2}}{\sqrt{(1-r)(\rho_+ r-
\rho_-)}}dr}=\frac{\int_0^1\frac{r^{3/2}}{\sqrt{(1-r)\Big(\frac{r}{\rho_+^2}-
\frac{\rho_-}{\rho_+^3}\Big)}}dr}{\int_0^1\frac{r^{1/2}}{\sqrt{(1-r)\Big(r-
\frac{\rho_-}{\rho_+}\Big)}}dr}.
\end{eqnarray*}
As we have seen the positive numerator is monotone increasing where the positive denominator is monotone decreasing so that the function $\mathcal{F}^0$ is monotone increasing. This finishes the proof of the lemma. \hfill $\square$
\begin{lemma}\label{func4}
For every $\gamma \neq 0$ we have $\lim_{x \to \infty}\mathcal{F}(\gamma,x)=0$.
\end{lemma}
\textbf{Proof:} Suppose that
$$x>\frac{2}{\gamma^2}$$
Then we can estimate
\begin{eqnarray*}
\sigma^+(\gamma,x)&=&2-\gamma^2 x+\sqrt{(2-\gamma^2x)^2+4(\gamma+1)^2}\\
&=&2-\gamma^2 x+(\gamma^2 x-2)\sqrt{1+\frac{4(\gamma+1)^2}{(\gamma^2 x-2)^2}}\\
&\leq&2-\gamma^2 x+(\gamma^2 x-2)\bigg(1+\frac{4(\gamma+1)^2}{2(\gamma^2 x-2)^2}\bigg)\\
&=&\frac{2(\gamma+1)^2}{\gamma^2 x-2}
\end{eqnarray*}
so that we have
$$\lim_{x \to \infty} \sigma^+(\gamma,x)=0.$$
The lemma follows. \hfill $\square$
\\ \\
Using these four lemma we are now in position to prove Theorem~\ref{main1}.
\\ \\
\textbf{Proof of Theorem~\ref{main1}: } Uniqueness is an immediate consequence of Lemma~\ref{func1}. To prove existence we note that from (\ref{func2}) and (\ref{func3}) we have for every $\gamma>0$
$$\mathcal{F}(\gamma,0) \geq \frac{8}{3}>2.$$
Since by Lemma~\ref{func4}
$$\lim_{x \to \infty} \mathcal{F}(\gamma,x)=0<2$$
the existence follows from the intermediate value theorem. \hfill $\square$

\section{Intersection}

In this section we examine for which values of $\mu$ the orbits $q_1$ and $q_2$ intersect and prove Theorem\,B from the introduction. 

\begin{lemma}\label{intlem1}
If $\kappa=0$, then $q^{\mathrm{max}}_1=\overline{q}_2$, i.e., the orbits of $q_1$ and $q_2$ touch each other, if
$\kappa>0$, then $q^{\mathrm{max}}_1>\overline{q}_2$, i.e., the orbits of $q_1$ and $q_2$ intersect, and if
$\kappa<0$, then $q^{\mathrm{max}}_1<\overline{q}_2$, i.e., the orbits of $q_1$ and $q_2$ do not intersect.
\end{lemma}
\textbf{Proof: } Combining (\ref{barqzwei2}) and (\ref{maximum}) we obtain
$$q_1^{\mathrm{max}}=\frac{\gamma^2 \kappa \overline{q}_1^2 + \sqrt{\gamma^4\kappa^2\overline{q}_1^4+4\overline{q}_2^2}}{2},$$
implying that
$$\bigg(q_1^{\mathrm{max}}-\frac{\gamma^2 \kappa \overline{q}_1^2}{2}\bigg)^2=\frac{\gamma^4 \kappa^2 \overline{q}_1^4}{4}+\overline{q}_2^2,$$
respectively,
$$\big(q_1^{\mathrm{max}}\big)^2=\overline{q}_2^2+\gamma^2 \kappa \overline{q}_1^2 q_1^{\mathrm{max}}.$$
The lemma is an immediate consequence of this formula. \hfill $\square$
\begin{lemma}\label{intlem2}
$\kappa=0$ happens if and only if $\gamma=\frac{3\pi-\varpi^2}{\varpi^2}$ where $\varpi$ is the lemniscatic constant. 
\end{lemma}
\textbf{Proof: } From (\ref{kapeq}) we see that $\kappa=0$ is equivalent to
$$1=\frac{2}{\gamma^2 \overline{q}_1}$$
or in other words
$$\overline{q}_1=\frac{2}{\gamma^2}.$$
Note that
$$\sigma^+\bigg(\gamma,\frac{2}{\gamma^2}\bigg)=2(\gamma+1)$$
respectively
$$\sigma^-\bigg(\gamma,\frac{2}{\gamma^2}\bigg)=-2(\gamma+1).$$
Hence (\ref{fixpunkt4}) becomes
$$
2=\frac{2(\gamma+1)\int_0^1\frac{r^{3/2}}{\sqrt{(1-r)(2(\gamma+1) r+2(\gamma+1))}}dr}{\int_0^1\frac{r^{1/2}}{\sqrt{(1-r)(2(\gamma+1) r+2(\gamma+1))}}dr}.
$$
which can be rewritten as
\begin{equation}\label{lemn1}
\frac{1}{\gamma+1}=\frac{\int_0^1\frac{r^{3/2}}{\sqrt{(1-r)(r+1)}}dr}{\int_0^1\frac{r^{1/2}}{\sqrt{(1-r)(r+1)}}dr}
=\frac{\int_0^1\frac{r^{3/2}}{\sqrt{1-r^2}}dr}{\int_0^1\frac{r^{1/2}}{\sqrt{1-r^2}}dr}
\end{equation}
Changing variables as
$$\sigma=r^2, \quad d\sigma=2r dr=2\sqrt{\sigma}dr$$
we obtain for the integral in the numerator
$$\int_0^1\frac{r^{3/2}}{\sqrt{1-r^2}}dr=\frac{1}{2}\int_0^1 \sigma^{1/4}(1-\sigma)^{-1/2}d\sigma=\frac{1}{2}
B\big(\tfrac{5}{4},\tfrac{1}{2}\big)=\frac{1}{2}\frac{\Gamma\big(\frac{5}{4}\big)\cdot \Gamma\big(\frac{1}{2}\big)}{\Gamma\big(\frac{7}{4}\big)}$$
where $B$ is the Beta function and $\Gamma$ is the Gamma function. For the denominator we get
$$\int_0^1\frac{r^{1/2}}{\sqrt{1-r^2}}d\rho=\frac{1}{2}\int_0^1 \sigma^{-1/4}(1-\sigma)^{-1/2}d\sigma=\frac{1}{2}
B\big(\tfrac{3}{4},\tfrac{1}{2}\big)=\frac{1}{2}\frac{\Gamma\big(\frac{3}{4}\big)\cdot \Gamma\big(\frac{1}{2}\big)}{\Gamma\big(\frac{5}{4}\big)}.$$
Plugging these formulas into (\ref{lemn1}) we have
\begin{equation}\label{lemn2}
\frac{1}{\gamma+1}
=\frac{\Gamma\big(\frac{5}{4}\big)^2}{\Gamma\big(\frac{7}{4}\big)\cdot \Gamma\big(\frac{3}{4}\big)}.
\end{equation}
Recall that the Gamma function satisfies
$$\Gamma(x+1)=x\Gamma(x).$$
Moreover, the Legendrian relation tells us
$$\Gamma\big(x\big)\Gamma\big(x+\tfrac{1}{2}\big)=2^{1-2x}\sqrt{\pi}\Gamma\big(x\big).$$
Using further
$$\Gamma\big(\tfrac{1}{2}\big)=\sqrt{\pi}$$
we compute
$$\Gamma\big(\tfrac{3}{4}\big)\Gamma\big(\tfrac{5}{4}\big)=2^{1-\frac{3}{2}}\sqrt{\pi}\Gamma \big(\tfrac{3}{2}\big)=
\frac{\sqrt{\pi}\Gamma\big(\frac{1}{2}\big)}{\sqrt{2}\cdot 2}=\frac{\pi}{\sqrt{2}\cdot 2}.$$
Taking advantage of
$$\Gamma\big(\tfrac{7}{4}\big)=\tfrac{3}{4}\Gamma\big(\tfrac{3}{4}\big)$$
we obtain from (\ref{lemn2})
\begin{eqnarray*}
\frac{1}{\gamma+1}&=&\frac{4}{3}\Bigg(\frac{\Gamma\big(\frac{5}{4}\big)}{\Gamma\big(\frac{3}{4}\big)}\Bigg)^2\\
&=&\frac{4}{3}\Bigg(\frac{\sqrt{2}\cdot 2 \Gamma\big(\frac{5}{4}\big)^2}{\pi}\Bigg)^2\\
&=&\frac{32}{3\pi^2}\cdot \Gamma\big(\tfrac{5}{4}\big)^4=\frac{32}{3\pi^2}\cdot \bigg(\frac{1}{4}\bigg)^4\cdot \Gamma\big(\tfrac{1}{4}\big)^4\\
&=&\frac{1}{24 \cdot \pi^2}
\Gamma\big(\tfrac{1}{4}\big)^4
\end{eqnarray*}
which with the help of the lemniscatic constant
$$\varpi=\frac{\Gamma\big(\frac{1}{4}\big)^2}{\sqrt{8\pi}}$$
becomes
$$\frac{1}{\gamma+1}=\frac{\varpi^2}{3\pi}.$$
Hence
$$\gamma=\frac{3 \pi}{\varpi^2}-1=\frac{3\pi-\varpi^2}{\varpi^2}.$$
This finishes the proof of the lemma. \hfill $\square$
\\ \\
Using (\ref{mueq}) we have the following immediate Corollary from the lemma.
\begin{cor}
$\kappa=0$ happens if and only if $\mu=\big(\tfrac{3\pi}{3\pi-\varpi^2}\big)^2 \cong 13,69$.
\end{cor}
\begin{lemma}\label{intlem3}
If $\gamma<\tfrac{3\pi-\varpi^2}{\varpi^2}$ the orbits $q_1$ and $q_2$ intersect and if $\gamma>\tfrac{3\pi-\varpi^2}{\varpi^2}$ they do not intersect. 
\end{lemma}
\textbf{Proof: } By definition of $\sigma^+$ we have
\begin{equation}\label{int1}
q_1^\mathrm{max}=\frac{\sigma^+}{2}\overline{q}_1.
\end{equation}
Using (\ref{fixpunkt4}) we have
\begin{equation}\label{int2}
\frac{2}{\sigma^+}=\frac{\int_0^1\frac{r^{3/2}}{\sqrt{(1-r)(\sigma^+ r-
\sigma^-)}}dr}{\int_0^1\frac{r^{1/2}}{\sqrt{(1-r)(\sigma^+ r-\sigma^-)}}dr}=
\frac{\int_0^1\frac{r^{3/2}}{\sqrt{(1-r)\big(r-
\frac{\sigma^-}{\sigma^+}\big)}}dr}{\int_0^1\frac{r^{1/2}}{\sqrt{(1-r)\big(r-\frac{\sigma^-}{\sigma^+}\big)}}dr}.
\end{equation}
Note that
$$a:=-\frac{\sigma^-}{\sigma^+}\geq 0.$$
Consider the function
$$a \mapsto \frac{\int_0^1\frac{r^{3/2}}{\sqrt{(1-r)(r-
a)}}dr}{\int_0^1\frac{r^{1/2}}{\sqrt{(1-r)(r-a)}}dr}.$$
This smooth function attains values in the open interval $(0,1)$ for all $a \geq 0$ and this is true even for its limit as $a$ goes to infinity. Therefore there exist 
$$0<c_-<c_+<1$$ such that for every $a \geq 0$
$$c_- \leq \frac{\int_0^1\frac{r^{3/2}}{\sqrt{(1-r)(r-
a)}}dr}{\int_0^1\frac{r^{1/2}}{\sqrt{(1-r)(r-a)}}dr}\leq c_+.$$
We conclude from (\ref{int1}) and (\ref{int2}) that
$$\tfrac{1}{c_-} \overline{q}_1 \leq q_1^\mathrm{max} \leq \tfrac{1}{c_+} \overline{q}_1.$$
Recall (\ref{barqzwei2}), namely
$$\overline{q}_2=(\gamma+1) \overline{q}_1.$$
Suppose that
$$\gamma<\tfrac{1}{c_-}-1.$$
Then
$$\overline{q}_2<\tfrac{1}{c_-}\overline{q}_1 \leq q_1^\mathrm{max}$$
which means that $q_1$ and $q_2$ intersect. By Lemma~\ref{intlem1} and Lemma~\ref{intlem2} we know that $q_1$ touches $q_2$ only if
$\gamma=\tfrac{3\pi-\varpi^2}{\varpi^2}$. Because $q_1$ and $q_2$ are uniquely determined by $\gamma$ in view of Theorem~\ref{main1} we conclude that for every $\gamma<\tfrac{3\pi-\varpi^2}{\varpi^2}$ the orbits $q_1$ and $q_2$ intersect.

Now suppose that
$$\gamma>\tfrac{1}{c_+}-1.$$
In this case
$$\overline{q}_2>\tfrac{1}{c_+}\overline{q}_1 \geq q_1^\mathrm{max}$$
so that the orbits $q_1$ and $q_2$ do not intersect. By the same reasoning as before we conclude that
for every $\gamma>\tfrac{3\pi-\varpi^2}{\varpi^2}$ the two orbits do not intersect. This finishes the proof of the lemma. \hfill $\square$
\\ \\
\textbf{Proof of Theorem\,B: } The theorem is an immediate consequence of Lemma~\ref{intlem3} by using $\mu=\tfrac{(\gamma+1)^2}{\gamma^2}$. \hfill $\square$
\\ \\
\emph{Acknowledgements: } The author acknowledges partial support by DFG grant FR 2637/2-2.

\end{document}